\documentclass[letterpaper, 10 pt, conference]{lib/ieeeconf}  

\IEEEoverridecommandlockouts                              

\usepackage{color}
\usepackage{algorithm}
\usepackage{algpseudocode}
\usepackage{amsmath}
\usepackage{fancyhdr}
\usepackage{amssymb}
\usepackage{type1cm}
\usepackage{url}
\usepackage{caption}
\usepackage[pdftex]{graphicx}
\usepackage{bm}
\usepackage{multirow}
\usepackage{subcaption}
\usepackage{tabularx}
\usepackage{ulem}
\usepackage{cite}

\usepackage{fancyhdr}
\title{\LARGE \bf
Decentralized Collective World Model \\for Emergent Communication and Coordination
}

\author{Kentaro Nomura$^{1}$, Tatsuya Aoki$^{1}$, Tadahiro Taniguchi$^{2,3}$ and Takato Horii$^{1,4}$
\thanks{*This work was supported by Japan Science and Technology Agency (JST) Moonshot R\&D Grant Number JPMJMS2011.}
\thanks{$^{1}$ Dept. of Systems Innovation, Graduate School of Engineering Science, The University of Osaka, Osaka, Japan} 
\thanks{\tt\small \{k.nomura@rlg., t.aoki@rlg., takato@\} sys.es.osaka-u.ac.jp}
\thanks{$^{2}$ Dept. of Informatics, Kyoto University, Kyoto, Japan}
\thanks{\tt\small taniguchi@i.kyoto-u.ac.jp}
\thanks{$^{3}$ Dept. of Science and Engineering, Ritsumeikan University, Shiga, Japan}
\thanks{$^{4}$ IRCN, The University of Tokyo, Tokyo, Japan}
}

\begin{document}

\maketitle
\thispagestyle{fancy}

\begin{abstract}

We propose a fully decentralized multi-agent world model that enables both symbol emergence for communication and coordinated behavior through temporal extension of collective predictive coding. Unlike previous research that focuses on either communication or coordination separately, our approach achieves both simultaneously. Our method integrates world models with communication channels, enabling agents to predict environmental dynamics, estimate states from partial observations, and share critical information through bidirectional message exchange with contrastive learning for message alignment. Using a two-agent trajectory drawing task, we demonstrate that our communication-based approach outperforms non-communicative models when agents have divergent perceptual capabilities, achieving the second-best coordination after centralized models. Importantly, our decentralized approach with constraints preventing direct access to other agents' internal states facilitates the emergence of more meaningful symbol systems that accurately reflect environmental states. These findings demonstrate the effectiveness of decentralized communication for supporting coordination while developing shared representations of the environment.

\end{abstract}

\section{Introduction}

Coordination through shared symbolic communication is fundamental to human society, enabling us to collectively achieve goals beyond individual capabilities \cite{sep-shared-agency}. 
As environments increasingly integrate artificial systems with humans, a critical challenge remains unsolved: how to enable distributed multi-agent systems to simultaneously develop shared symbol systems and effective coordination without centralized control. This paper addresses this challenge by proposing an approach that integrates world models with communication channels, enabling agent groups to form symbol systems while coordinating through communication in partially observable environments. Our method allows agents to use world models to estimate environmental states from partial observations, share this information through emergent communication via bidirectional message exchange using the acquired common symbol system, and achieve coordination in complex, partially observable settings.

Coordination in multi-agent systems requires a shared symbol system where all participants interpret symbols consistently \cite{DBLP:journals/corr/abs-2012-08630}. 
Communication plays a crucial role by allowing agents to exchange independently acquired information and infer environmental states \cite{wang2021tomc}, enabling them to complement each other's knowledge and abilities, especially in partially observable environments \cite{DBLP:journals/corr/abs-2012-08630,DBLP:journals/corr/abs-2006-02419}.

\begin{figure}[t]
    \centering
    \includegraphics[width=1.0\linewidth]{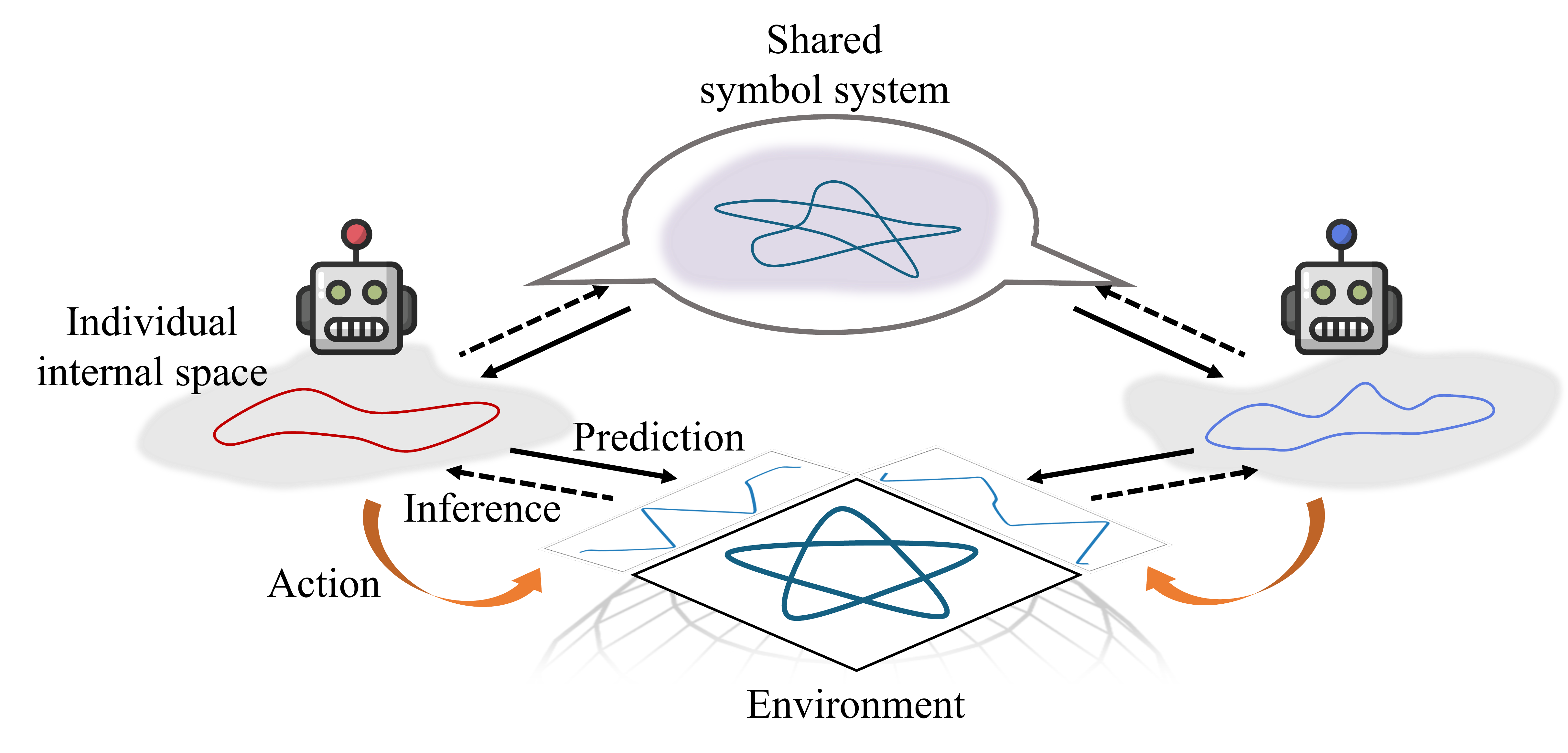}
    \caption{Overview of the proposed method. Each agent perceives a partial region of the environment, while complementing knowledge of other regions through communication. This leads to the emergence of a symbol system that represents collective knowledge.}
    \label{fig:overview}
\end{figure}

Current emergent communication approaches fail to address the dual challenge of symbol system learning and coordination in dynamic environments. Research in emergent communication has primarily focused on language development through games like Signaling Games and Referential Games \cite{DBLP:journals/corr/abs-2006-02419,brandizzi2023}. 
However, these studies are typically limited to one-way communication and do not address multi-step action determination necessary for coordination in dynamic environments. 

World models-based approaches involve agents maintaining internal models that learn environmental dynamics and state estimation, forming latent state space representations from raw sensorimotor information \cite{NEURIPS2018_2de5d166,hafner2024masteringdiversedomainsworld, Taniguchi03072023worldmodeldevrob}. These learned representations enable agents to predict future states and infer hidden environmental information from incomplete observations, potentially enhancing coordination capabilities in multi-agent settings \cite{Peters2025}.
However, current world model approaches in multi-agent settings lack mechanisms for collective knowledge formation through distributed symbol systems.
Traditional multi-agent reinforcement learning methods often employ parameter sharing or centralized learning \cite{NIPS2017_68a97503,wen2022}, becoming computationally expensive and impractical in real-world settings \cite{DBLP:journals/corr/abs-2110-14555}.

Decentralized approaches to multi-agent coordination lack mechanisms for developing shared symbol systems. 
Recent research has explored decentralized approaches that transmit information through non-differentiable messages \cite{wang2025learning,NEURIPS2021_80fee67c,Pina2024}, making them suitable for real-world applications where centralized control is impractical \cite{nowak1999}. 
However, existing decentralized methods typically focus on immediate task achievement rather than developing shared symbol systems that capture environmental dynamics. 
This limitation restricts their ability to form a common understanding that would enable more sophisticated coordination in complex, changing scenarios.

Existing symbol emergence methods based on Collective Predictive Coding (CPC) \cite{taniguchi2024cpch,taniguchi2024generativeemergentcommunicationlarge} are limited to static observations and do not capture environmental dynamics. CPC extends predictive coding \cite{Rao1999} and the free energy principle (FEP) \cite{Friston2010} to social domains, viewing language and symbols as collective knowledge formed through the distributed participation of individual agents . Various CPC-based approaches have been developed, including methods based on Markov Chain Monte Carlo \cite{okumura2023mhng,Taniguchi02102023} and contrastive learning \cite{hoang2024simsiamnaminggameunified}. Some research has applied these concepts to multi-agent coordination by inferring symbols for multi-step action selection \cite{nakamura2023cai,ebara2023}. However, existing methods primarily focus on forming symbol systems from static observations at specific moments, making them inadequate for environments with continuous changes and temporal dependencies.

Our proposed approach integrates world models with communication channels to enable simultaneous symbol system formation and coordination in dynamic environments. As illustrated in Figure \ref{fig:overview}, we integrate world models with communication channels, allowing agents to predict environmental dynamics and share critical information through emergent communication. By reinterpreting FEP-based formulation of CPC \cite{taniguchi2024collectivepredictivecodingmodel} in a fully distributed form, we develop a system where agents learn to communicate without centralized control. Our contrastive learning mechanism aligns messages across agents, creating a unified symbol system that emerges naturally through learning.

This work makes three significant contributions to multi-agent coordination and communication research. First, we implement a two-agent world model that integrates CPC-based symbol emergence with temporal dynamics learning in a fully decentralized manner. Second, we demonstrate that environment-general symbols learned through world model prediction—rather than task-specific communication protocols—can effectively support coordination in partially observable environments, allowing agents to complement each other's limited perceptions. Third, we show that the distributed constraints of our approach lead to the formation of meaningful symbol systems that represent the global environmental state, emerging naturally from predictive learning of environmental dynamics. From the perspective of CPC research, our method extends previous CPC-based approaches temporally, providing a framework that captures environmental dynamics rather than static observations. 
While our approach shares similarities with recent work by \cite{lo2024learning} in using contrastive learning for message alignment, we uniquely develop environment-general symbolic representations through predictive modeling.

\section{Formalization of Multi-Agent Interaction \\in Partially Observable Environments}
In single-agent scenarios, the interaction between an agent and its environment is classically formalized as a Markov Decision Process (MDP). However, real-world agents rarely observe environmental states directly, instead perceiving their surroundings through limited sensory observations. To account for this limitation, world models and FEP approaches model such interactions as Partially Observable Markov Decision Processes (POMDPs).

When multiple agents interact within a shared environment, the dynamics become more complex. For fully observable multi-agent settings, the interaction is formalized as a Markov Game \cite{LITTMAN1994157}. A Markov Game with $K$ agents consists of a tuple $\langle \mathcal{K}, \mathcal{S}, \{\mathcal{A}^k\}_k, P \rangle$, where $\mathcal{K}$ represents the set of agents, $\mathcal{S}$ denotes the state space, and $\mathcal{A}^k$ specifies the action space for agent $k$. The transition function $P(s_{t+1}\mid s_t, \{a^k_t\}_k)$ determines state transitions based on the current state $s_t$ and actions $a^k_t \in \mathcal{A}^k$ taken by all agents.

In decentralized systems, agents operate autonomously without centralized control, making environmental state information inaccessible to individual agents. Moreover, each agent receives distinct observations based on its unique perspective. Such partially observable, distributed settings are formalized through Decentralized Partially Observable MDPs (Dec-POMDPs) \cite{bernstein2002decpomdp}.
A Dec-POMDP is structured as a tuple $\langle \mathcal{K}, \mathcal{S}, \{\mathcal{A}^k\}_k, P, \{\mathcal{O}^k\}_k, \{\Omega^k\}_k \rangle$, where $\mathcal{O}^k$ represents agent $k$'s private observation space, and $\Omega^k(o^k_t\mid s_t)$ denotes the observation function governing how agent $k$ perceives state $s_t$.

Our approach aims to model cooperative agent groups within Dec-POMDP environments, where a fundamental challenge emerges: environmental states evolve as a function of all agents' collective actions. This collective influence means that each agent's observations are affected not only by its own actions but also by those of other agents in the system. Consequently, from any individual agent's perspective, the environment exhibits non-stationary characteristics, which significantly complicates predictive modeling.

The partial observability inherent to Dec-POMDP presents an additional challenge for optimal decision-making. When an agent's sensory input captures only a subset of the environmental state, its ability to comprehend the global environment is fundamentally limited. 

To address this limitation, we employ CPC, which introduces shared symbolic representations functioning as distributed knowledge across all agents. This framework enables the integration of diverse agent perspectives through message exchange. 
Our method leverages this emergent property of CPC to enable coordinated action determination by synthesizing: (1) partial observation obtained through interaction with environment, and (2) messages from other agents. This dual-channel approach enhances decision-making quality while maintaining the decentralized nature of the system.

\section{Proposed model}


In this section, we first describe the centralized model architecture and objective function for our two-agent system. Next, we explain how this centralized formulation is decomposed for decentralized implementation through approximation methods. Finally, we detail the action determination procedure of agents.

\subsection{CPC-based Multi-agent World Model}
\label{mawm}
\begin{figure}[t]
    \centering
    \includegraphics[width=1.0\linewidth]{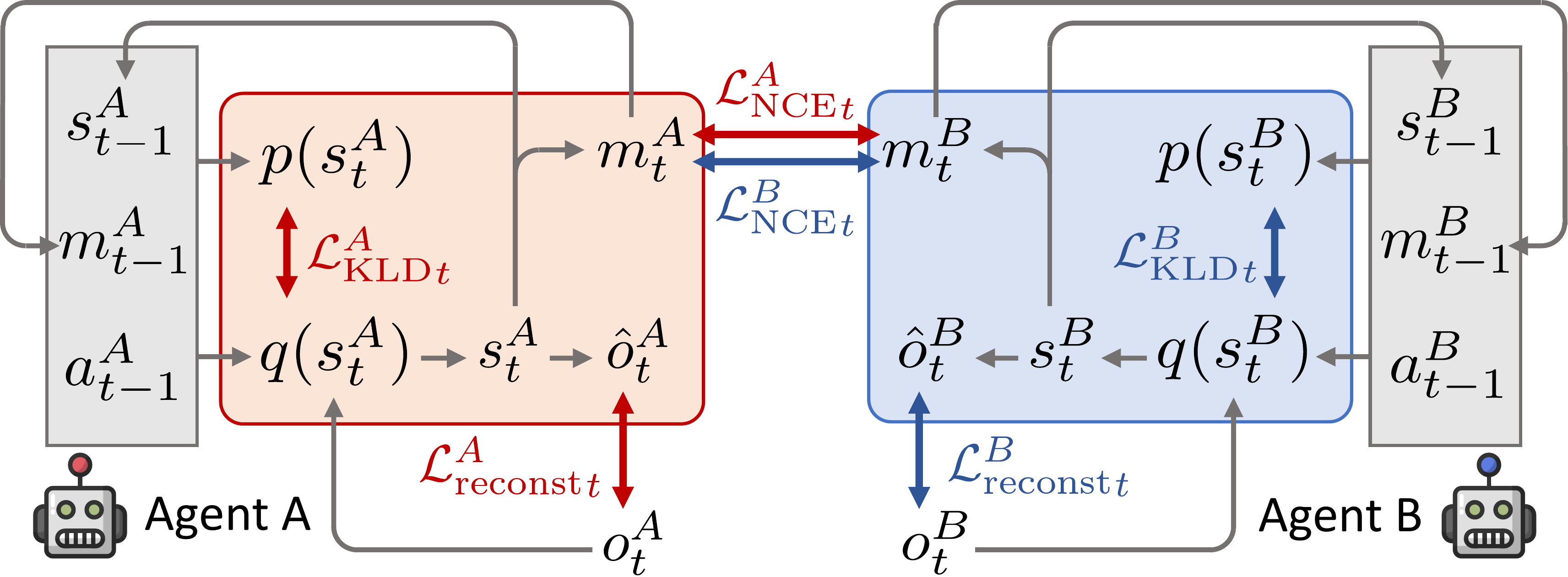}
    \caption{The architecture of the collective world model. Bidirectional arrows indicate losses for training.}
    \label{fig:train}
\end{figure}
Following the CPC framework, we first formulate a centralized model that assumes access to shared messages between agents, then decompose it for decentralized implementation. Our proposed model, illustrated in Figure \ref{fig:train}, is constructed around interconnected agent world models that jointly infer messages through exchange.
Each world model learns environmental dynamics as state transitions in a latent space by modeling the observation generation process based on a POMDP. We employ the Recurrent State Space Model \cite{hafner2018planet} as the foundational architecture for world models. 
Our key innovation is the introduction of a probabilistic variable $m_t$ representing messages exchanged between agents, which follows a continuous distribution enabling effectively infinite possible messages, and influences the generation of each agent's latent variables at each time step.
For a system with two agents ($\mathrm{A}, \mathrm{B}$), the multi-agent world model comprises the following components:
\begin{align}
\label{eq:ma_rssm_components}
\begin{split}
    &\text{Message inference model:}\quad\quad q(m_t\mid s_t^\mathrm{A}, s_t^\mathrm{B}),\\
    &\text{Representation model:}~ q(s_t^*\mid s_{t-1}^*, m_{t-1}, a_{t-1}^*, o_t^*),\\
    &\text{Message generation model:}\quad p(m_t),\\
    &\text{Observation model:}\quad\quad\quad\quad p(o_t^*\mid s_t^*),\\
    &\text{Transition model:}\quad p(s_t^*\mid s_{t-1}^*, m_{t-1}, a_{t-1}^*),\\  
    &\quad\quad\quad\quad\quad\quad\quad\quad\quad\quad\quad\quad\quad\quad\quad
     \mathrm{for}~* \in \{\mathrm{A}, \mathrm{B}\}.
\end{split}
\end{align}
Here, $o_t^*,~a_t^*$, and $s_t^*$ represent the observation, action, and latent variable of agent $*$ at time $t$, respectively. The latent state $s^*_t$ encompasses both a probabilistic latent variable $z^*_t$ and a deterministic latent variable $h^*_t$, where $h_t^*$ is the internal state of the Gated Recurrent Unit \cite{cho2014learning} in agent $*$'s world model. Additionally, $m_t$ denotes a probabilistic variable representing the common message shared between agents at time $t$.

The multi-agent world model is trained by minimizing the variational free energy (VFE), which establishes an upper bound on the negative log-likelihood. To distinguish between the VFE of the entire group and that of each individual agent, we designate the former as the collective free energy (CFE). The CFE is formulated as:
\begin{align}
    \label{eq:cfe}
    F &= \sum_{t=1}^T \Bigg[\sum_{*\in\{\mathrm{A},\mathrm{B}\}}\bigg\{ - \mathbb{E}_{q}\left[\log p({o}_t^*\mid {s}_t^*)\right] \notag\\
    &\quad+ D_{KL}\Big[q(s_t^*\mid s_{t-1}^*, m_{t-1}, a_{t-1}^*, o_t^*)||\notag\\
    &\quad\quad\quad\quad\quad p(s_t^*\mid s_{t-1}^*, m_{t-1}, a_{t-1}^*)\Big]\bigg\}  \notag\\
    &\quad+ D_{KL}\left[q(m_{t-1}\mid s_{t-1}^\mathrm{A}, s_{t-1}^\mathrm{B})||p(m_{t-1})\right]\Bigg]\notag\\
    &= \sum_{t=1}^T \Bigg[\sum_{*\in\{\mathrm{A},\mathrm{B}\}} \underbrace{{\mathcal{L}^*_{\mathrm{reconst}}}_t + {\mathcal{L}^*_{\mathrm{KLD}}}_t}_{\text{Individual VFE}} \notag\\
    &\quad+ \underbrace{D_{KL}\left[q(m_{t-1}\mid s_{t-1}^\mathrm{A}, s_{t-1}^\mathrm{B})||p(m_{t-1})\right]}_{\text{Collective regularization term}}\Bigg].
\end{align}
The first and second terms represent the individual VFE and have the effect of forming internal representations. The third term is the collective regularization (CR) term, which promotes message formation \cite{taniguchi2024collectivepredictivecodingmodel}.

\subsection{Approximation of the CR term for Distributed Learning}
A significant challenge emerges in implementing the aforementioned model. While our theoretical formulation presupposes the existence of a model $q(m_t\mid s_t^\mathrm{A}, s_t^\mathrm{B})$ that infers messages from the latent variables $s_t^*$ of two agents, such a model cannot be directly computed in a decentralized system. In natural symbol emergence, an individual's cognitive system is confined to its sensorimotor boundaries, rendering it impossible for an agent to directly access other agents' internal representations. This constraint makes the CR term incalculable in its original form, presenting a substantial impediment to implementing distributed learning across agents.

To address this constraint, we propose an approximation methodology that enables independent learning for each agent. We introduce independent probabilistic variables $m^\mathrm{A}_t$ and $m^\mathrm{B}_t$ representing each agent's estimate of the shared message $m_t$. We then approximate both the prior distribution $p(m_t)$ and posterior distribution $q(m_t\mid s_t^\mathrm{A}, s_t^\mathrm{B})$ of the message using a Product-of-Experts (PoE) formulation:
\begin{align}
    p(m_t) &\approx C_{\text{pm}}\prod_{*\in \{\mathrm{A},\mathrm{B}\}} p(m^*_t),\\
    q(m_t\mid s_t^\mathrm{A}, s_t^\mathrm{B}) &\approx C_{\text{qm}}\prod_{*\in \{\mathrm{A},\mathrm{B}\}} q(m^*_t\mid s_t^*)
\end{align}
where $C_{\text{pm}}$ and $C_{\text{qm}}$ denote normalization constants.
This approximation is theoretically justified for both posterior and prior distributions in our context. For posterior distributions, this approach aligns with realistic constraints where agents can only infer messages based on their own internal representations—precisely matching the decentralized nature of the learning problem. For prior distributions, the approximation is appropriate when interpreting collective prior knowledge about symbols as an integration of individual agents' knowledge bases. This formulation not only maintains computational tractability but also naturally promotes consensus in message representation, which is fundamental for symbol emergence in multi-agent systems.

With this approximation established, we must determine how to set the prior distribution of messages $p(m^*_t)$ for each agent. We propose defining $p(m^*_t)$ as a distribution that reflects the collective prior knowledge at that time. Specifically, we consider a distribution based on what messages the other agent is inferring at that moment. Thus, we define each agent's prior distribution as the other agent's posterior distribution:
\begin{align}
    \label{eq:m_prior}
    p(m^\mathrm{A}_t) \triangleq q(m^\mathrm{B}_t\mid {s}_t^\mathrm{B}),\quad p(m^\mathrm{B}_t) &\triangleq q(m^\mathrm{A}_t\mid {s}_t^\mathrm{A}).
\end{align}
This formulation establishes a feedback mechanism wherein each agent's message inference is influenced by the collective understanding of the other agent, facilitating convergence to a shared communication protocol.

 We designate the agent that infers messages from observations and transmits them to the other agent as the Speaker, and the agent that updates parameters based on observations and received messages from the other agent as the Listener. The Listener's CR term at time $t$ is the KL divergence related to both agents' message posterior distributions:
\begin{align}
    \label{eq:m_kl}
    &D_{KL}\left[q(m^\mathrm{Sp}_t\mid {s}_t^\mathrm{Sp})||p(m^\mathrm{Sp}_t)\right]\notag\\
    &\quad =D_{KL}\left[q(m^\mathrm{Sp}_t\mid {s}_t^\mathrm{Sp})||q(m^\mathrm{Li}_t\mid {s}_t^\mathrm{Li})\right]
\end{align}
where $s^\mathrm{Li}_t,~m^\mathrm{Li}_t$ and $s^\mathrm{Sp}_t,~m^\mathrm{Sp}_t$ represent the latent variables and messages of the Listener and Speaker, respectively. A practical challenge arises in calculating equation \eqref{eq:m_kl}, as it requires the parameters of the message posterior distributions. However, in realistic scenarios, agents communicate by exchanging sampled messages rather than distributional parameters. Each agent samples messages from its posterior distribution and exchanges these samples with other agents. Consequently, an agent has access only to message samples from other agents, not to the parameters of their posterior distributions. To address this limitation, we leverage the Noise Contrastive Estimation framework \cite{gutmann2010nce} and employ the InfoNCE loss \cite{oord2019representationlearningcontrastivepredictive} to estimate the message inference model from samples alone:
\begin{align}
    &{\mathcal{L}^\mathrm{Li}_{\mathrm{NCE}}}_t \notag\\
    &= -\mathbb{E}_{\substack{q(m^\mathrm{Sp}_t\mid {s}_t^\mathrm{Sp}),\\q(m^\mathrm{Li}_t\mid {s}_t^\mathrm{Li})}}\left[\log\frac{\mathrm{sim}(m^\mathrm{Li}_t,~m^\mathrm{Sp}_t)}{\mathbb{E}_{q({m'}^\mathrm{Sp}_t\mid {s'_t}^\mathrm{Sp})}\left[\mathrm{sim}(m^\mathrm{Li}_t,~{m'}^\mathrm{Sp}_t)\right]}\right].
\end{align}

In practice, each agent $*\in \{\mathrm{A},\mathrm{B}\}$ learns by minimizing the weighted distributed CFE $F^*$:
\begin{align}
    \label{eq:cfe_final}
    F^* &= \sum_{t=1}^T ({\mathcal{L}^*_{\mathrm{reconst}}}_t + w_{\mathrm{KLD}}{\mathcal{L}^*_{\mathrm{KLD}}}_t + w_{\mathrm{NCE}}{\mathcal{L}^*_{\mathrm{NCE}}}_t).
\end{align}
Since $F^*$ can be computed independently for agents $\mathrm{A}$ and $\mathrm{B}$, we optimize $F^\mathrm{A}$ and $F^\mathrm{B}$ simultaneously during training, enabling decentralized multi-agent world model learning.

\subsection{Action Determination through Communication}\label{subsec:action-determination}
In Dec-POMDPs, effective cooperation requires agents to infer the complete environmental state despite having only partial observations. This limitation highlights the critical importance of information exchange among agents to achieve comprehensive environmental understanding.

Agents generate actions by inferring internal representations and messages based on observation history. To mitigate computational challenges with increasing temporal horizons, we implement a sliding time window of length $W$, wherein information beyond $W$ timesteps is disregarded. Each agent maintains queues of maximum length $W+1$ for observations and $W$ for past actions.

\begin{algorithm}[t]
\caption{Action Determination}
\label{alg:action decision}
\begin{algorithmic}[1]
\Procedure{Action Determination}{$o^\mathrm{A}_{t-W:t}$, $a^\mathrm{A}_{t-W:t-1}$, $s^\mathrm{A}_{t-W}$, $o^\mathrm{B}_{t-W:t}$, $a^\mathrm{B}_{t-W:t-1}$, $s^\mathrm{B}_{t-W}$}
    \State $m^\mathrm{A}_{t-W:t-1} \sim q(\cdot\mid s^\mathrm{A}_{t-W}, a^\mathrm{A}_{t-W:t-1}, o^\mathrm{A}_{t-W:t})$
    \State $m^\mathrm{B}_{t-W:t-1} \sim q(\cdot\mid s^\mathrm{B}_{t-W}, a^\mathrm{B}_{t-W:t-1}, o^\mathrm{B}_{t-W:t})$ \Comment{(1)}
    \If{$\mathcal{L}^\mathrm{A}_{\text{VFE}}(m^\mathrm{A}_{t-W:t-1}) \leq \mathcal{L}^\mathrm{A}_{\text{VFE}}(m^\mathrm{B}_{t-W:t-1})$} \hspace{-4.5pt}\Comment{(2,3)}
        \State $\hat{a}^\mathrm{A}_t \gets \pi^\mathrm{A}(s^\mathrm{A}_t, m^\mathrm{A}_{t-1})$
    \Else
        \State $\hat{a}^\mathrm{A}_t \gets \pi^\mathrm{A}(s^\mathrm{A}_t, m^\mathrm{B}_{t-1})$  \Comment{(4)}
    \EndIf
    \If{$\mathcal{L}^\mathrm{B}_{\text{VFE}}(m^\mathrm{B}_{t-W:t-1}) \leq \mathcal{L}^\mathrm{B}_{\text{VFE}}(m^\mathrm{A}_{t-W:t-1})$} \hspace{-4.5pt}\Comment{(2,3)}
        \State $\hat{a}^\mathrm{B}_t \gets \pi^\mathrm{B}(s^\mathrm{B}_t, m^\mathrm{B}_{t-1})$
    \Else
        \State $\hat{a}^\mathrm{B}_t \gets \pi^\mathrm{B}(s^\mathrm{B}_t, m^\mathrm{A}_{t-1})$  \Comment{(4)}
    \EndIf
    \State \Return $(\hat{a}^\mathrm{A}_t, \hat{a}^\mathrm{B}_t)$
\EndProcedure
\end{algorithmic}
\end{algorithm}

When determining actions, agents infer internal representations and messages using the information stored in their respective queues. These internal representations are influenced not only by the agent's observations and actions but also by messages received in the previous timestep. For optimal coordination, agents must utilize messages that maximize observation predictability.
Our model enables agents to infer messages within a shared representational space and compare self-generated messages with those received from others. Each agent selects the message that minimizes the VFE, thereby inferring representations that best explain past observations.
Each agent determines actions according to the following steps (Algorithm \ref{alg:action decision}):

\begin{enumerate}
    \item {\bf Inference of Internal Representations and Messages}:
    Agents receive observations $o^*_t$ from the environment, add them to their queues, and perform sequential inference of internal representations and messages based on their observation-action sequences.
    \begin{align}
        &s^*_{t-\tau}\sim q^*(\cdot|s^*_{t-\tau-1},m^*_{t-\tau-1},a^*_{t-\tau-1},o^*_{t-\tau}),\\
        &m^*_{t-\tau}\sim q^*(\cdot|s^*_{t-\tau})\quad\textrm{for}~\tau=W,\ldots,1.
    \end{align}
    \item {\bf Message Exchange}: Both agents reciprocally transmit their inferred message sequences $m^*_{t-W:t-1}$ that were derived in step 1.
    \item {\bf Selection of Messages and Internal Representations}:
    Each agent reconstructs observations using both self-generated and received messages, computes the individual VFE for each, and selects the message sequence yielding the minimum VFE.
    \item {\bf Action Determination}: 
    Each agent determines its action using its current internal representation $s^*_t$ and the terminal message $m^*_{t-1}$ from the selected message sequence. The determined action is then pushed into the agent's action queue.
\end{enumerate}

\section{experiment}

\subsection{Task Setup}
We conducted experiments in a simulated environment to investigate whether multiple agents can achieve coordinated behavior through message formation that represents the complete environmental state.
The experimental task involved two agents collaboratively moving a point $P$ in two-dimensional space to trace a predefined trajectory. 
Each agent receives sensory signals from point P coordinates through agent-specific sensory modules.
These modules discretize one coordinate axis into a predetermined number of bins and introduce noise. Specifically, agent A's sensory module discretizes the y-axis coordinates, while agent B's module discretizes the x-axis coordinates.
By manipulating the number of bins in these sensory modules, we can systematically control the extent of environmental information accessible to each agent. With an infinite number of bins, no discretization occurs, enabling complete environmental perception. Conversely, with a single bin (bin=1), agents can observe only one axis, effectively reducing each agent's perceptible environmental state by half. This configuration allows us to transition from a Markov Game setting (infinite bins) to a Decentralized Partially Observable Markov Decision Process (Dec-POMDP) setting (finite bins).
\begin{figure}[t]
    \centering
    \begin{tabular}{cc}
        \begin{minipage}[t]{0.47\hsize}
            \centering
            \includegraphics[width=0.814\linewidth]{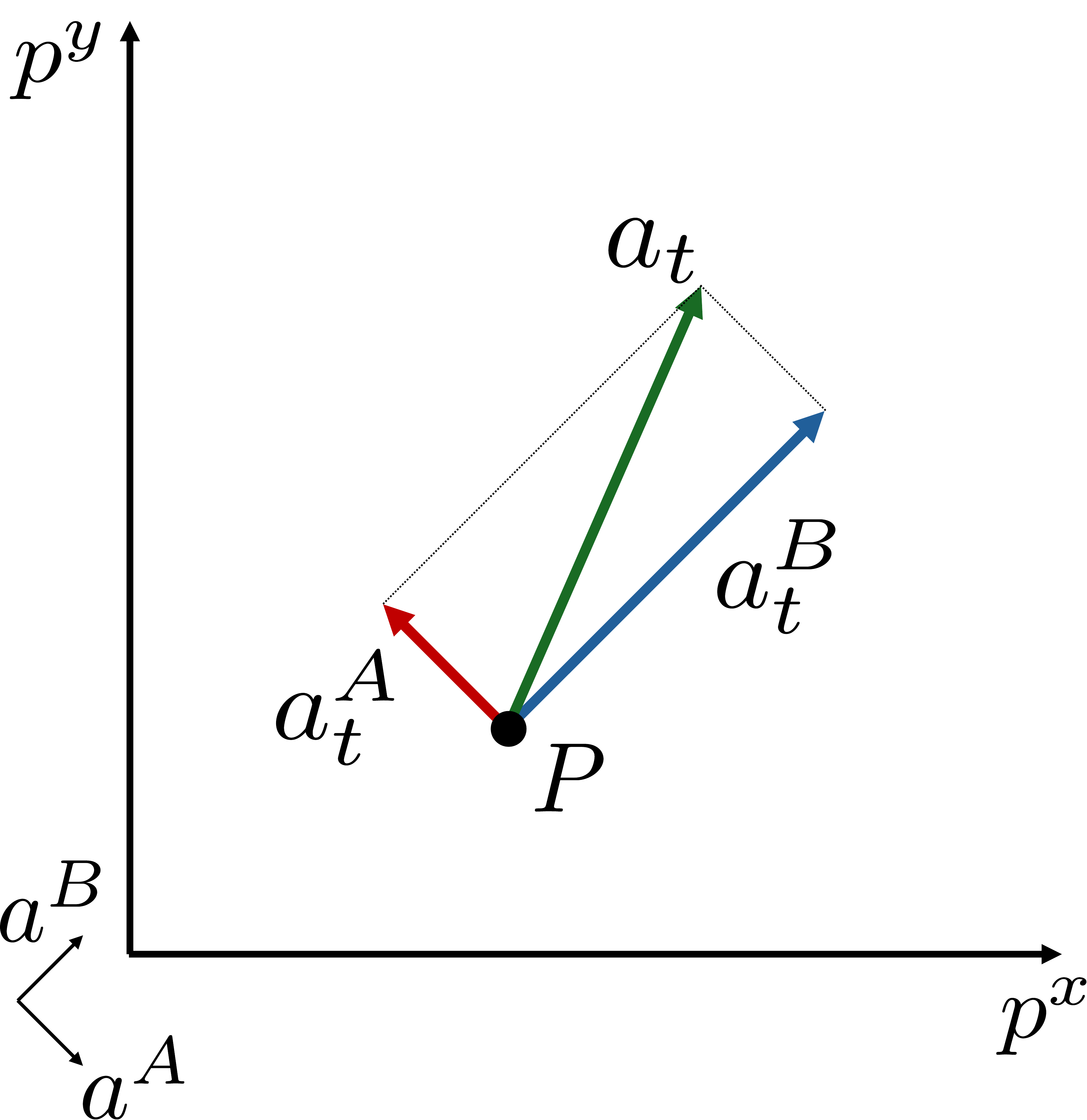}
            \vspace{-10pt}
            \caption*{(a)}
        \end{minipage} &\hspace{-20pt}
        \begin{minipage}[t]{0.5\hsize}
            \centering
            \includegraphics[width=0.82\linewidth]{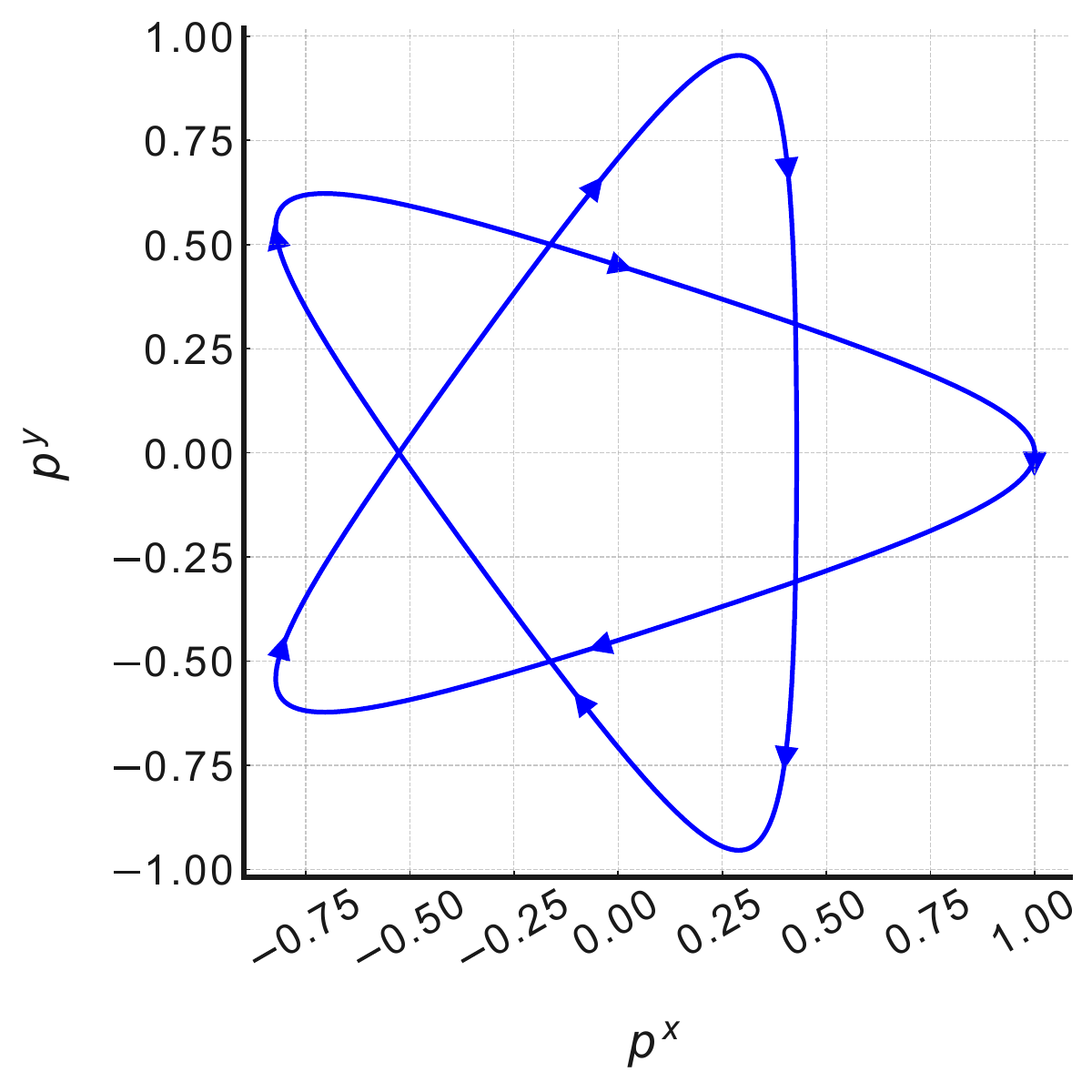}
            \vspace{-10pt}
            \caption*{(b)}
        \end{minipage}
    \end{tabular}
    \caption{(a) Schematic overview of the trajectory drawing coordination task environment created for the experiment. (b) The hypotrochoid trajectory that agents were required to draw in this experiment.}
    \label{fig:exp_setting}
    \vspace{-0mm}
\end{figure}

Each agent controls the velocity along one of the axes in a coordinate system derived by rotating the x-y coordinate system by $-\pi / 4$. This rotation ensures that the fully observable axis coordinates are affected by both agents' actions. Figure \ref{fig:exp_setting} illustrates the environmental configuration, depicting both observation and action axes.
We implemented a task requiring agents to draw a star-shaped hypotrochoid curve. 

We established five distinct configurations for the number of bins in each agent's sensory module: infinite, 8, 6, 2, and 1. For each configuration, we generated 2000 samples of coordinated expert data, with each sample completing one circuit of the trajectory in 200 steps.
We examine how the agents' capacity to develop effective symbolic representations of the environment influences their coordination capabilities, particularly as their individual observational capabilities become increasingly constrained.

\subsection{Model Architecture and Training Setup}
Both agent models were constructed with identical architectures, although their parameters were independently initialized with different random values. To enhance learning stability, we implemented the distribution of the latent variable $z^*_t$ as a unimix categorical distribution \cite{hafner2024masteringdiversedomainsworld} with a 1\% uniform mixture component, utilizing a 4-dimensional representation with 4 distinct classes.
For the message distribution, we employed a two-dimensional multivariate Gaussian distribution. The internal state dimension of the GRU, $h^*_t$, was set to 32. Within the CFE (equation \eqref{eq:cfe_final}), the weighting coefficients $w_{\textrm{KLD}}$ and $w_{\textrm{NCE}}$ were set to 0.01 and 0.005, respectively. For the InfoNCE loss, we utilized negative squared Euclidean distance as the similarity function, incorporating a temperature parameter $\tau$ set to 2.0:
\begin{align}
    \textrm{sim}(m, m') = - \frac{\|m-m'\|^2_2}{\tau}.
\end{align}

In this experiment, we trained the policy through behavioral cloning of expert demonstrations. Specifically, we defined the policy learning objective as the Mean Squared Error between the predicted actions and the expert actions:
\begin{align}
    \mathcal{L}^*_\pi = \sum_{t}\|\pi^*(s^*_t, m^*_{t-1}) - a^*_t\|^2_2.
\end{align}

All models were trained using a batch size of 500 for 1,000 epochs to ensure convergence of the learning process.

\subsection{Conditions}
To systematically investigate the effects of decentralization and symbol emergence through the introduction of InfoNCE loss on world model learning and cooperative behavior acquisition, we established three experimental conditions:

\begin{itemize}
    \item {\bf EmergentCommunication (EC):} Our proposed method, in which agents develop symbol systems with InfoNCE loss while constructing decentralized world models.
    \item {\bf BrainConnected (BC):} A condition without InfoNCE loss and without model decentralization. In this setup, message inference distribution is conditioned on both agents' internal states, wherein the message-inferring model incorporates both agents' latent states $s^A_t,~s^B_t$ as inputs. The loss function comprises the sum of individual VFEs for each agent.
    \item {\bf NoCommunication (NC):} A condition where both agents maintain and train independent world models. The internal representations $s^*_t$ and messages $m^*_t$ of each agent remain independent, with each agent optimizing its Individual VFE as the loss function.
\end{itemize}

Additionally, we established a control condition featuring a single agent with complete environmental observability as a performance baseline.

For all experimental conditions, we conducted training and evaluation across three distinct random seeds. For each seed, we generated 100 test samples for comprehensive evaluation. The time window parameter $W$ used for EC evaluation was fixed at 10 throughout our experiments.

\section{Result and Discussion}

\subsection{The Effectiveness of Decentralized Symbol Emergence}
\begin{figure}[t]
    \centering
    \includegraphics[width=1.0\linewidth]{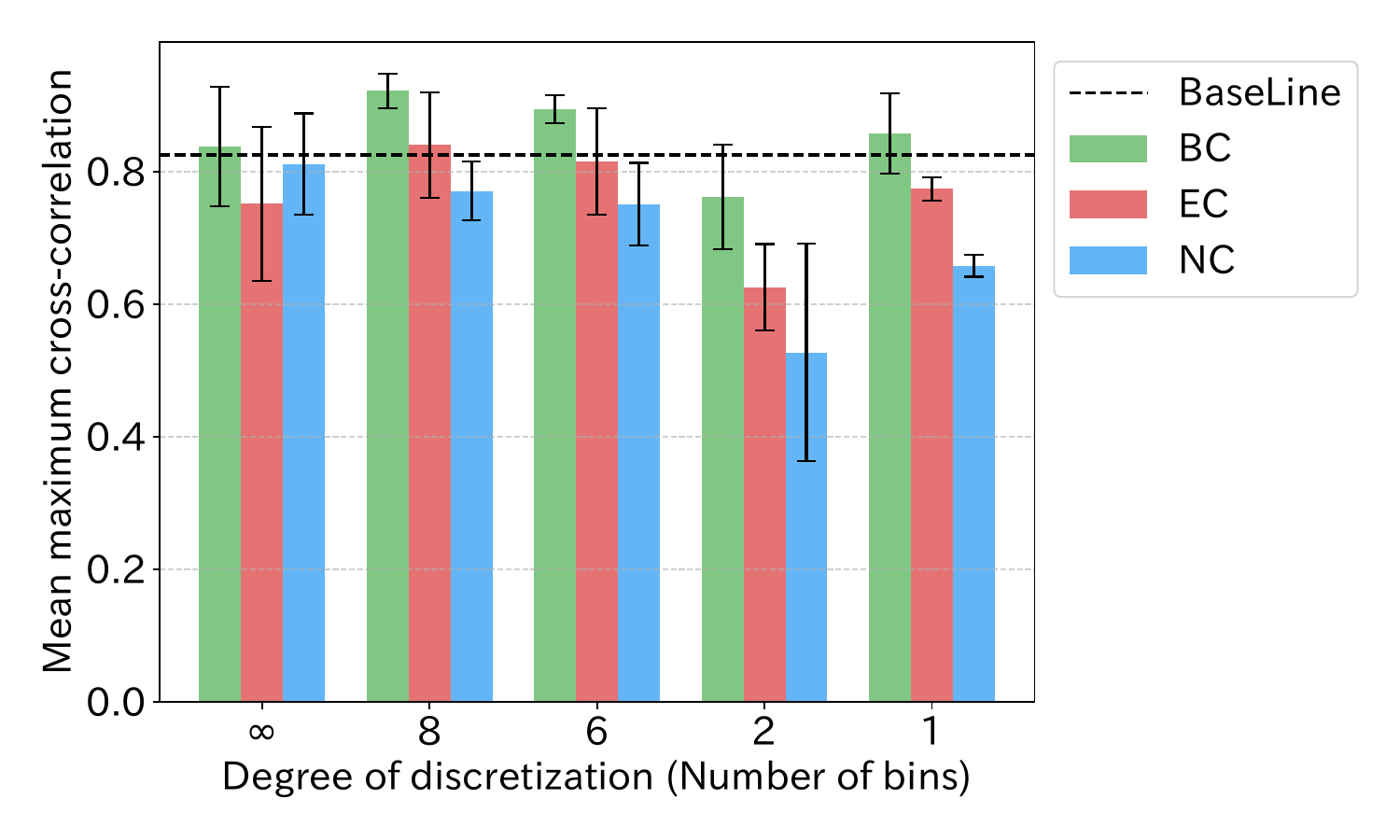}
    \caption{Comparison of coordination achievement across learning conditions. The values represent the average of maximum cross-correlation between trajectories drawn by agents and test data. Error bars represent standard deviation.}
    \label{fig:ccr_bar_cdt}
\end{figure}

We compared coordination performance across learning conditions. Point $P$ was randomly positioned at the beginning of each trial, with 100 trials conducted using the learned policies to generate trajectories. Coordination success was quantified using the maximum cross-correlation between the generated trajectories and the ideal hypotrochoid trajectory.

Figure \ref{fig:ccr_bar_cdt} presents the mean scores for each condition and bin configuration. In the Markov Game condition (infinite number of bins), agents had complete access to the environmental state, enabling them to select optimal actions without requiring information exchange with other agents. This resulted in minimal performance differences between conditions, with all approaches achieving scores comparable to the baseline.
Conversely, in Dec-POMDP conditions (finite bin configurations), the BC condition demonstrated superior coordination, followed by the EC condition, while the NC condition exhibited the lowest performance. The enhanced performance of both EC and BC conditions, which leverage shared messages between agents, compared to the NC condition where agents learn independently, clearly demonstrates that communication channels built upon unified representational systems facilitate more effective coordination.

Examining the performance gap between EC and NC conditions, we observe that as the number of observation bins decreases, this gap widens significantly. This finding suggests that communication through shared symbol systems becomes increasingly critical for successful coordination when agents experience greater disparities in information accessibility—such as when they operate at considerable physical distances or when they possess different observable information modalities.

\subsection{The Effectiveness of Symbolic Communication}
\begin{figure}
    \centering
    \includegraphics[width=1.0\linewidth]{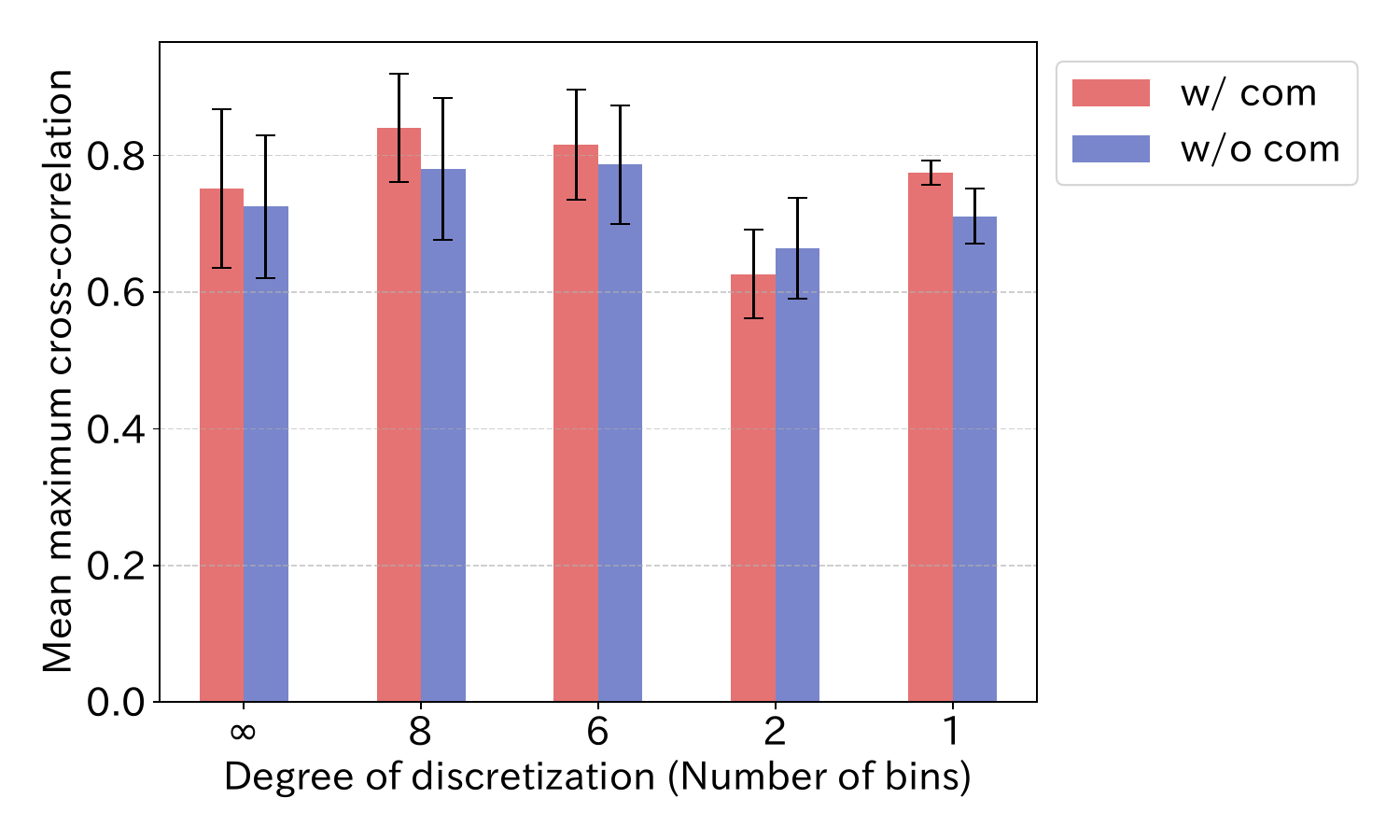}
    \caption{Comparison of coordination achievement with (\texttt{w/ com}) and without (\texttt{w/o com}) communication through message exchange using the EC (proposed method) model. Error bars represent standard deviation.}
    \label{fig:ccr_bar_com}
\end{figure}

We investigated whether symbolic communication through message exchange, as described in Section \ref{subsec:action-determination}, enhances coordination success. We compared performance between two conditions: \texttt{w/ com}, which employs our proposed method with communication, and \texttt{w/o com}, where agents make decisions using only self-inferred messages without communication.

Figure \ref{fig:ccr_bar_com} illustrates these results. For conditions with bin numbers exceeding 1, \texttt{w/ com} exhibited higher performance than \texttt{w/o com} in most cases, although with substantial overlap in standard deviation ranges, indicating minimal differences.
Notably, when the bin number equals 1—representing scenarios where observable information differs completely between agents—\texttt{w/ com} consistently outperformed \texttt{w/o com}.
These results provide compelling evidence that symbolic communication offers particular advantages when agents possess heterogeneous perceptual capabilities.

\subsection{Analysis of Emergent Messages}
We examined how effectively the message structure formed through learning reflected the environmental state. Using Representational Similarity Analysis (RSA) \cite{kriegeskorte2008rsa}, we evaluated the structural similarity between messages inferred during test data reconstruction and the trajectory of point $P$.
For each episode, we calculated distances between data points at all possible time step combinations for both inferred messages and actual point $P$ coordinates, creating dissimilarity matrices. We then extracted the upper triangular components (excluding diagonal elements) and calculated Spearman's rank correlation. Higher scores indicate greater structural similarity between the point $P$ trajectory and the sequence of inferred messages, demonstrating that the model's acquired messages more accurately represent the environmental state.  This metric corresponds to the concept of Positive Signaling \cite{lowe2019}, which measures message effectiveness between agents by quantifying the correlation between an agent's inferred messages and its observations.

\begin{figure}[t]
    \centering
    \includegraphics[width=1.0\linewidth]{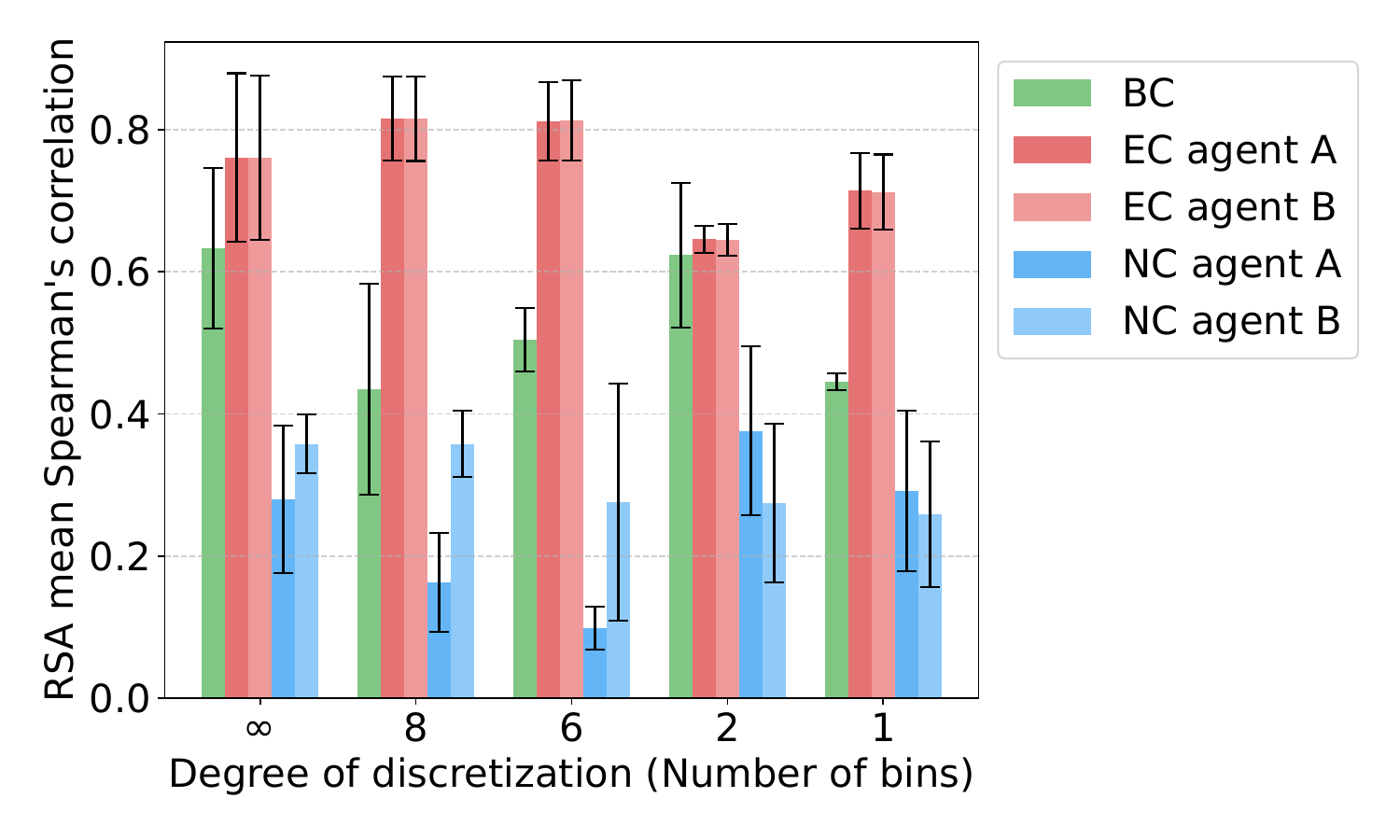}
    \caption{Similarity between the structure of inferred messages when reconstructing test data observations and the structure of the actual trajectory of point $P$, as calculated by RSA. Error bars represent standard deviation.}
    \label{fig:rsa}
\end{figure}

\begin{figure}[t]
    \centering
    \includegraphics[width=1.0\linewidth]{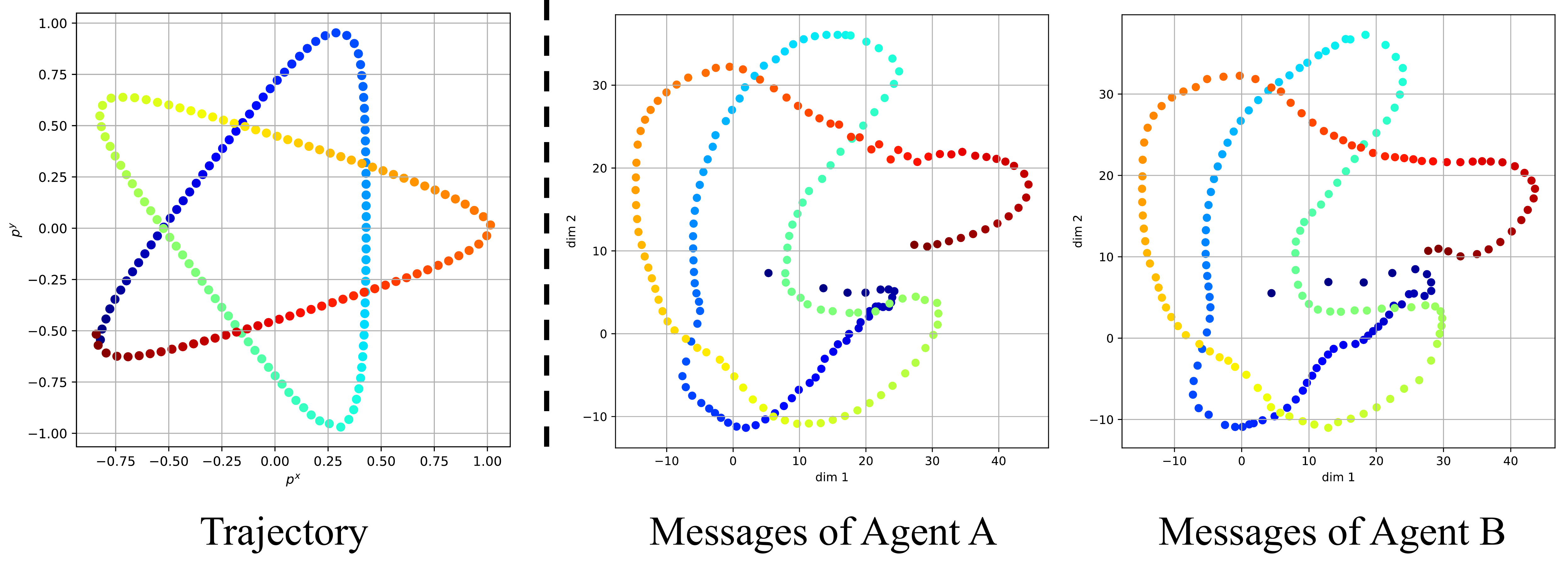}
    \caption{(Left) Trajectory of point $P$ when moved according to test data, and (Right) sequence of messages inferred by each agent when reconstructing observations using EC (proposed method) with 6 bins. In all plots, the color of points changes from blue to red as time steps progress.}
    \label{fig:messages}
\end{figure}

Figure \ref{fig:rsa} presents a comparative analysis of scores across learning conditions. It should be noted that for the BC condition, only one message is inferred since the models of each agent are connected through message variables. Across all conditions, messages acquired through EC demonstrated superior environmental state representation compared to other conditions. Significantly, EC outperformed BC despite the latter allowing agents direct access to other agents' internal states. This finding suggests that in CPC, distributed constraints—where agents cannot directly access other agents' internal states or observations—actually facilitate the emergence of more meaningful symbol systems.

Figure \ref{fig:messages} illustrates the trajectory of inferred messages in the EC, showcasing a trial example with 6 bins. Comparison with the actual trajectory of point $P$ reveals that the actual environmental state is effectively captured in the message representation space, despite each agent receiving only partial observations.
Furthermore, due to the alignment effect of InfoNCE loss, nearly identical messages are inferred by different agents at each time step, despite variations in their observations and actions.

Although EC acquired meaningful message representations, its coordination performance remained below that of BC as previously noted. This discrepancy likely stems from limitations in our policy learning methodology, where the representations acquired by world models were not optimally utilized by the policies. Our experiments employed only imitation learning of expert actions without incorporating exploratory behaviors or reinforcement learning. Without experiencing coordination failures during training, agents lacked the critical negative feedback necessary to refine their coordination strategies, which likely contributed to the observed performance gap between EC and BC.

\section{conclusion}
We proposed a method for multiple agents to acquire a common symbol system for mutual communication and achieve coordinative behavior in Dec-POMDP settings. Our multi-agent world model incorporates a communication channel enabling information exchange through a shared representational system. Based on CPC, we derived a learning rule operating in a completely distributed manner and proposed an action determination algorithm where agents interact through the acquired communication channel.

We designed a task requiring two agents to cooperatively trace a desired trajectory and compared our approach with centralized and non-communicative models. The results confirmed that symbol system formation and communication contribute to coordination, particularly when observable information differs between agents. We demonstrated that meaningful messages effectively reflecting the overall environmental state can emerge through distributed learning.

Future work includes incorporating reinforcement learning and active inference into policy learning, extending our approach to larger agent groups, and evaluating performance on more complex tasks with higher-dimensional observations. In this work, we focus on message emergence that supports cooperation, while exploration of environments by multi-agents and skill acquisition from scratch will be addressed in future work.








\bibliographystyle{lib/IEEEtran.bst}

\end{document}